# $Ni_2X_2$ (X=pnictide, chalcogenide, or B) Based Superconductors

F. Ronning[1], E.D. Bauer[1], T. Park[1,2], N. Kurita[1], T. Klimczuk[1,3], R. Movshovich[1], A.S. Sefat[4], D. Mandrus[4], J.D. Thompson[1]

[1] *Los Alamos National Laboratory. Los Alamos, NM 87545 USA*

[2] *Department of Physics, Sungkyunkwan University, Suwon 440-746, Korea*

[3] *Faculty of Applied Physics and Mathematics, Gdansk University of Technology, Narutowicza 11/12, 80-952 Gdansk, Poland*

[4] *Material Science & Technology Division, Oak Ridge National Laboratory, Oak Ridge, TN 37831, USA*

**Abstract**

We review the properties of Ni-based superconductors which contain $Ni_2X_2$ (X=As, P, Bi, Si, Ge, B) planes, a common structural element found also in the recently discovered FeAs superconductors. Strong evidence for the fully gapped nature of the superconducting state has come from field dependent thermal conductivity results on $BaNi_2As_2$. Coupled with the lack of magnetism, the majority of evidence suggests that the Ni-based compounds are conventional electron-phonon mediated superconductors. However, the increase in $T_c$ in LaNiAsO with doping is anomalous, and mimics the behavior in LaFeAsO. Furthermore, comparisons of the properties of Ni- and Fe-based systems show many similarities, particularly with regards to structure-property relationships. This suggests a deeper connection between the physics of the FeAs superconductors and the related Ni-based systems which deserves further investigation.

**Introduction**:

Even before superconductivity was found in LaFeAs(O,F) at 26 K[1] it was known that the Ni-analog LaNiPO superconducts at 4.3 K[2,3]. While many compounds containing $Fe_2As_2$ planes with transition temperatures well above 5 K have been found (e.g. [4,5,6,7]), to date, none of the nickel analogs (excluding the nickel borocarbides) have $T_c$s exceeding 5K[2,3,8,9,10,11,12,13,14,15,16,17,18]. There are two possible reasons for this: (1) superconductivity in the nickel compounds has no relation to the iron-based systems and is likely a conventional phonon mediated BCS s-wave superconductor, or (2) the pairing mechanism is in fact the same as in the iron-based systems, but the conditions for superconductivity are not nearly as optimized for the Ni-based systems as they are for the Fe-based systems. At this time, no definitive conclusion concerning which scenario is correct can be made.

In this review, we first examine the properties of Ni-based superconductors which contain the $Ni_2X_2$ (X = As, P, Bi, Si, Ge, B) PbO-type structure. We attempt to identify common features and then examine various compound specific studies and/or properties. In the second half of the review, we attempt to compare the Ni systems to the Fe analogs. Specifically, we compare doping studies, band structure calculations, and structural relationships. When examined independently, the majority of evidence suggests that the Ni-based systems are simple conventional BCS superconductors. However, the similarity of several relationships among the Fe- and Ni-based compounds suggests deeper connections between the two compounds. We conclude with some open questions specific to Ni-based compounds.

**Properties of Ni-based compounds**:

The fact that many Ni-analogs of the FeAs based superconductors also superconduct suggests that a detailed comparison of these systems may reveal how $T_c$ is enhanced by more than an order of magnitude in the Fe-based compounds relative to the Ni-based compounds. Of course, one should not presume that structural similarity guarantees similar physics. Certainly, $(La,Ba)_2CuO_4$[19], $Sr_2RuO_4$[20], and $(Ba,K)BiO_3$[21] all have the perovskite structure, but each is believed to have entirely different mechanisms for superconductivity. However, the extreme robustness of superconductivity to structural variations, elemental substitution, and doping for both the Fe- and Ni-based compounds offers a unique opportunity in the study of superconductivity regardless of the pairing mechanism. Consequently, we begin by reviewing the properties of the Ni-based superconductors which contain puckered $Ni_2X_2$ planes, common in the higher temperature Fe-based systems, with limited consideration of their Fe counterparts. For the moment we exclude the nickel-borocarbides[22,23] and -boronitrides[24] from our discussion, despite them also possessing this structural element as shown in figure 1g. Implicit in our discussion will be the blind assumption that all these Ni-based systems have the same superconducting pairing mechanism.

Figure 1 displays the crystal structures that contain the checkerboard $Ni_2X_2$ planes and have been found to be superconducting. The structure types include: ZrCuSiAs, $ThCr_2Si_2$, $Pr_3Cu_4P_4O_2$, and $U_3Ni_4Si_4$. Table 1 lists the known superconducting Ni-based compounds along with their properties. The most significant observation is that none of the compounds display superconducting transition temperatures above 5 K. The superconducting upper critical field $H_{c2}$ is rather small, with the exception of ~5 T in doped LaNiAsO and 1.2 T in $La_3Ni_4Si_4$. The density of states at the Fermi level is small in all compounds, with Sommerfeld coefficients $\gamma$ ranging from 4.35 mJ/(mol-Ni $K^2$) in $SrNi_2As_2$ to 9.23 mJ/(mol-Ni $K^2$) in $La_3Ni_4Si_4$. In cases where $H_{c2}$ is small, specific heat or magnetization measurements reveal that the systems can have very small Ginzburg-Landau parameters $\kappa$ (near the border between type I and type II), most notably κ = 2.1 in SrNi$_2$P$_2$. The specific heat jump at T$_c$ is often less than the weak coupling BCS value of ΔC/ γT$_c$ = 1.43. This is rather peculiar, as X-ray measurements exclude the possibility of structurally unique phases to less than 5% in these cases, which would suggest a more exotic interpretation. However, it is still possible that the reduced jump is due to impurities, either through pair breaking effects[25], or by an impurity "phase" which is structurally similar, but differs more subtly, for example, through site substitution[26] or nickel vacancies[27].

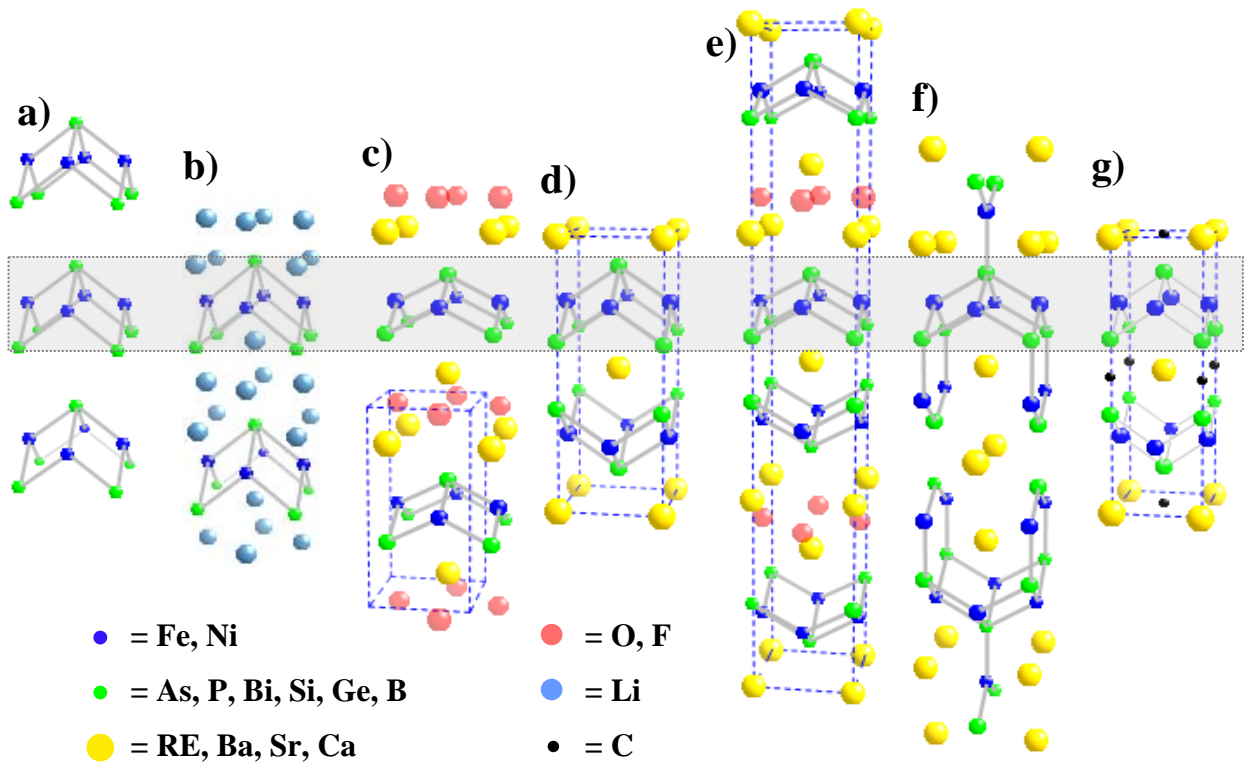

Figure 1. (color online) The crystal structure-types [a) PbO, b) Cu$_2$Sb, c) ZrCuSiAs, d) ThCr$_2$Si$_2$, e) Pr$_3$Cu$_4$P$_4$O$_2$, f) U$_3$Ni$_4$Si$_4$, g) YNi$_2$B$_2$C] which support superconductivity for either T = Fe (e.g. α-FeSe, LiFeAs, LaFeAs(O,F), (Ba,K)Fe$_2$As$_2$ for (a), (b), (c), and (d) respectively) and/or T = Ni (See table 1 for compounds with structures of (c), (d), (e), and (f)). All have in common the T$_2$X$_2$ structural element highlighted in the figure. RE = Rare Earth. (g) illustrates the same structural element in the borocarbide structure.

Table 1: Properties of Ni-based superconductors which possess a common $Ni_2X_2$ structural unit. Samples are polycrystaline unless otherwise noted.

| | $T_c$ (K) | $H_{c2}(0)$ (T) | $H_{c1}$ (Oe) | $\gamma_{exp}$ (mJ/mol -Ni $K^2$) | $\gamma_{th}$ (mJ/mol -Ni $K^2$) | $\lambda$ | $\Theta_D$ (K) [f] | $\Delta C/\gamma T$ | refs |
|---|---|---|---|---|---|---|---|---|---|
| LaNiPO [s] | 4.3[b,e] | ~0.2[d,e] | 25@1.8K[e] | | 3.32 | | | | 2,3,28 |
| $La_3Ni_4P_4O_2$ | 2.2[f] | 0.59[d] | 520[e] | 6.22 | | 0.5[a] | 357 | 1.25 | 8 |
| $BaNi_2P_2$ [s] | 2.7[d] | 0.16[gc], 0.065[hc] | 50@2K[eg], 80@2K[ec] | | 4.66 | | | | 9,29,30 |
| $SrNi_2P_2$ [s] | 1.4[f] | 0.039[f,g,h] | 88[f] | 7.5 | 3.72[i] | 1.02[j] | 348 | 1.27 | 10,29 |
| LaNiAsO | 2.75[b] | ~0.2[e] | 15@1.8K[e] | | 3.81 | | | | 11,36 |
| LaNiAs(O,F) | 3.8[b,f] | 4.6[b] | | 7.3 | | 0.92[j,k] | | 1.9 | 12 |
| (La,Sr)NiAsO | 3.7[b] | 5.5[b] | | | | | | | 13 |
| $BaNi_2As_2$ [s] | 0.68[f] | 0.11[f,g] | 5[f] | 6.15 | 4.19[i] | 0.47[j] | 206 | 1.31 | 14,31,32 |
| $SrNi_2As_2$ [s] | 0.62[f] | 0.015[g,c], 0.021[h,c] | | 4.35 | | | 244 | | 15 |
| $LaNiBiO_{1-x}$ | 4.25[b] | ~3[b] | | | 5.39 | | | | 16,33 |
| GdNiBiO | 4.5[b] | 2.5[b] | | | | | | | 17 |
| $La_3Ni_4Si_4$ | 1.0[f] | 1.2[b] | 107[e] | 9.23 | | 0.4[a] | 321 | 1.32 | 18,34 |
| $La_3Ni_4Ge_4$ | 0.7[f] | 0.26[b] | 68[e] | 8.63 | | 0.4[a] | 256 | 0.95 | 18,34 |

[a] = from McMillan formula[35]
[b] = determined by resistive onset
[c] = determined by resistive midpoint
[d] = determined by $\rho=0$
[e] = determined by magnetization measurements
[f] = determined by $C_p$
[g] = H//ab
[h] = H//c
[i] = does not fully account for the structural transformation from the $ThCr_2Si_2$ structure
[j] = ($\lambda = \gamma_{exp}/\gamma_{th} - 1$)
[k] = use $\gamma_{th}$ from LaNiAsO
[s] = single crystal

A major question is whether any competing phases such as magnetism can be found in the Ni-based systems. Early calculations suggested that both the Ni- and Fe-compounds are in close proximity to a magnetic instability[36]. A few systems show structural transitions, such as $BaNi_2As_2$ and $SrNi_2P_2$. By analogy with $AFe_2As_2$ (A=Ba, Sr, Eu, Ca)[37,38,39,40,41], one can speculate whether the structural transition coincides with a magnetic transition. However, to date there is no evidence for magnetism (by which we mean long range order, evidence for magnetic moments on the Ni-sites, and/or fluctuations indicating proximity to magnetic order) in any of the Ni-based systems. The one exception is $La_3Ni_4P_4O_2$ which displays an enhanced Wilson ratio[8], and therefore, may be close to a ferromagnetic instability. Preliminary neutron scattering measurements[42] have placed a weak upper limit of $1\mu_B$ in $BaNi_2As_2$ based on the lack of **Q** dependence to the observed nuclear Bragg peaks. More importantly, $^{31}P$ NMR failed to observe any magnetism in $SrNi_2P_2$[10]. Consequently, it appears that magnetism is very weak (if at all present) in the Ni-based systems.

### System specific studies/properties:

#### $BaNi_2As_2$

Single crystals of $BaNi_2As_2$ have been grown by both Pb flux[14] and NiAs flux[43]. The superconducting transition temperature as well as the structural transition temperature is independent of the growth technique, indicating that site substitution by the Pb flux, as occurs in $BaFe_2As_2$[26], does not occur here. $BaNi_2As_2$ has a structural transition at 130 K that causes very strong twinning. It has been identified as a tetragonal to triclinic structural transition[43], and contrary to other structural transitions in the $ThCr_2Si_2$ structure, it is relatively insensitive to pressure and there is no enhancement of superconductivity up to 2.5 GPa[44]. Due to the similarity of the structural transition to the ones found in $AFe_2As_2$ (A = Ba, Sr, Ca, Eu), it was suggested that magnetism may also be involved in the transition. However, as mentioned above, to date there has been no evidence for magnetism in any of the Ni-based systems.

Bulk superconductivity in $BaNi_2As_2$ at $T_c = 0.68$ K is well established by heat capacity, AC magnetic susceptibility, thermal conductivity, and resistivity. From the thermal conductivity data and band structure calculations a mean free path $\ell = 70$Å is estimated, which places

BaNi$_2$As$_2$ in the dirty limit ($\ell/\xi < 1$) when compared with the coherence length $\xi = 550$ Å estimated from H$_{c2}$. The heat capacity data in figure 2a can be integrated and the change in free energy equated to the thermodynamic critical field. This gives a thermodynamic critical field H$_c$ = 73.1 Oe. Thus, from the relationships $\kappa = H_{c2}/\sqrt{2}H_c = H_c/\sqrt{2}H_{c1} = \lambda/\xi$, one finds BaNi$_2$As$_2$ is a type II superconductor with $\kappa = 11$, H$_{c1} = 5$ Oe and $\lambda = 6000$ Å.

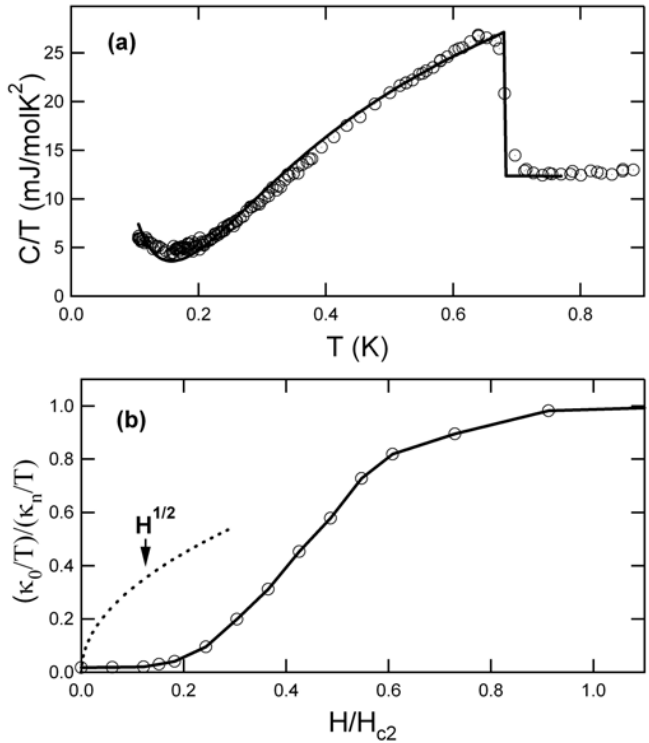

Figure 2. (a) Heat capacity data of BaNi$_2$As$_2$ which is well fit (solid line) by an s-wave BCS expression plus a low temperature nuclear Schottky term (discussed in ref [31]). (b) Field dependence of the residual linear term $\kappa_0/T$ of the thermal conductivity normalized by the normal state value above H$_{c2}$. The concave dependence is anticipated for fully gapped superconductors, while convex behavior is expected for superconductors with nodes in the superconducting order parameter. Specifically, $\sqrt{H}$ is expected for line nodes as shown by the dashed line.

As shown in figure 2, the heat capacity data is well fit by a BCS s-wave expression, and more importantly the thermal conductivity shows concave field dependence of the residual linear term $\kappa_0/T$. This is strong evidence that BaNi$_2$As$_2$ is a fully gapped superconductor[31], as nodal planes would create an easily distinguishable convex field dependence to the thermal conductivity (e.g. $\kappa_0/T \propto \sqrt{H}$ for line nodes). Furthermore, as can be seen in Figure 4, the Fermi surface is sufficiently complex to rule out the possibility that the nodal planes (for example of an s± state) simply fail to intersect the Fermi surface, as has been argued to occur in the Fe-based systems.

### $EuNi_2As_2$

While both $BaNi_2As_2$ and $SrNi_2As_2$ are superconducting near 0.65 K, from resistivity data (not shown)[45] $EuNi_2As_2$ is not superconducting down to 0.03 K despite $Eu^{2+}$ having a similar ionic size to $Sr^{2+}$. In $EuFe_2As_2$, the Eu moments order antiferromagnetically at 20 K[46]. With K doping on the Eu site both the magnetic ordering of the Eu moment, and the AF ordering of the Fe moments is suppressed, and superconductivity at 32 K is observed[47]. (The value of $T_c$ is similar to that of K-doped $SrFe_2As_2$.) However, with Ni doping on the Fe site the SDW transition associated with the magnetic moments on the Fe sites can be fully suppressed, but the magnetic moments on the Eu site remain ordered at roughly 18 K (although they switch to ferromagnetic order) and no superconductivity is observed down to 2K[48]. It appears that the ordering of the $Eu^{2+}$ moments prevents the occurrence of superconductivity in the doped $EuFe_2As_2$ compound. This is also the case for the Ni-analog. In $EuNi_2As_2$ the onset of antiferromagnetic order of the Eu moments at 14 K[49,50] is too strong for superconductivity to occur in the $Ni_2As_2$ sublattice down to 30mK (a factor of 20 times smaller than $T_c$ for $SrNi_2As_2$).

### $SrNi_2P_2$

$SrNi_2P_2$ is unique among the Ni-based systems in the $ThCr_2Si_2$ structure as it also undergoes a structural transition into the so-called collapsed tetragonal phase at relatively low pressures[51,52]. At room temperature this occurs at 4kbar. CaFe$_2$As$_2$ possesses a similar structural transition under pressure[53], and depending on the pressure transmitting medium used, one does[54,55] or does not[56] find evidence for superconductivity. At ambient pressure SrNi$_2$P$_2$ undergoes a structural phase transition at 325 K from a high temperature tetragonal to low temperature orthorhombic structure, which may be thought of as a precursor to the collapsed tetragonal phase. At low temperature, some of the P atoms in neighboring planes possess a strongly bonded configuration, whereas they are all weakly bonded to each other at high temperatures. Superconductivity is found at $T_c$ = 1.4 K at ambient pressure. With applied pressure, all the P atoms adopt a strongly bonded configuration as the system enters the "collapsed tetragonal" phase which has the same symmetry (but smaller volume and presumably more 3D) as the high temperature tetragonal phase at ambient pressure. Superconductivity is still observed in this collapsed tetragonal state, although the transition temperature is substantially reduced compared to the ambient pressure result[10] (e.g. $T_c$ = 0.6 K at 7.3 kbar). This supports the notion that reduced dimensionality is better for superconductivity.

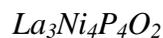

*La$_3$Ni$_4$P$_4$O$_2$*

La$_3$Ni$_4$P$_4$O$_2$ is a particularly interesting compound in the family of Ni-based superconductors. The structure is that of alternating layers of LaNiPO and LaNi$_2$P$_2$. The former is a superconductor with $T_c$ = 4.3 K, while the latter does not superconduct down to 1.8 K. One would anticipate that the dimensionality of La$_3$Ni$_4$P$_4$O$_2$ lies between LaNiPO (more 2-D) and LaNi$_2$P$_2$ (more 3-D), and with a $T_c$ = 2.2 K La$_3$Ni$_4$P$_4$O$_2$ could be used to support the argument that reduced dimensionality is a means for achieving higher transition temperatures. However, initial band structure calculations contradict the naïve assumption that La$_3$Ni$_4$P$_4$O$_2$ is electronically more 3-D than LaNiPO[57]. Thus, more work is clearly needed to understand the relation of superconductivity in the $(LaNi_2P_2)_m(LaNiPO)_n$ family of compounds.

## Doping Studies

Relatively little has been done with regards to chemical substitution in Ni-based systems. However, the effects of doping on LaNiAsO are remarkable. Both hole doping with Sr replacing La[13], and electron doping with F replacing O[12] *increases* the superconducting transition temperature as shown in figure 3. Furthermore, $H_{c2}$ is also increased by a factor of nearly 20 in the doped system compared to the LaNiAsO parent compound. Interestingly, the pressure dependence of $T_c$ is non-monotonic for both LaNiPO and LaNiAsO[58], an unusual behavior for conventional superconductors below 25kbar and resembling that of LaFeAs(O,F)[59]. Sr doping into GdNiBiO also has a slight increase in $T_c$[17]. This dependence on doping is highly unusual. Furthermore, there is no indication from band structure calculations (such as the chemical potential of the parent LaNiAsO compound lying at a local minimum in the density of states) to anticipate such an effect. More work is needed to understand this behavior. For completeness, we mention that several substitution studies have also been performed on $SrNi_2P_2$, but only the influence on the structural transition was investigated[60].

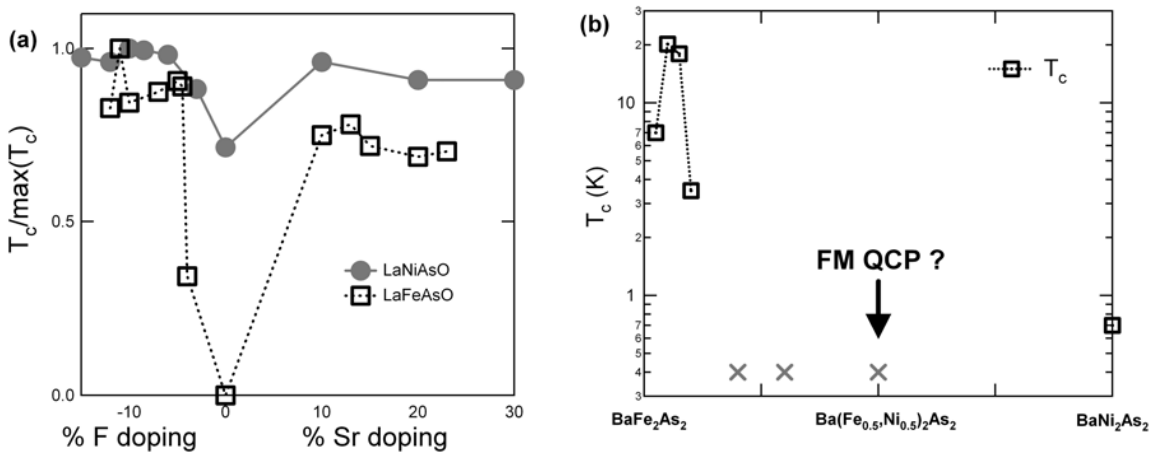

Figure 3. (a) Doping phase diagram of (La,Sr)TAs(O,F) (T = Ni or Fe). Values reproduced from refs [1,12,13,61]. (b) Doping phase diagram of Ba(Fe,Ni)$_2$As$_2$. × - indicate lowest temperature measured without observation of superconductivity. Values reproduced from refs [14,43,62]

**Electronic Structure**

With Ni having 2 additional electrons relative to Fe, the band structures of Ni-based compounds[16,28,30,32,33,36,57,63,64,65] are reasonably approximated by a rigid band shift relative to the Fe analogs. A consequence of this is that the Fermi surfaces of these compounds can differ rather markedly, and the carrier density is much larger in the Ni-analogs[66]. The small pockets which existed in the Fe compound now are larger sheets crossing through much of the Brillouin zone (see figure 4), which has important consequences for several proposed order parameter symmetries such as the s± state.

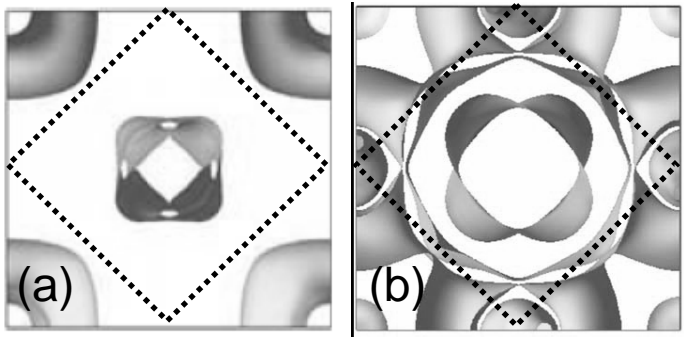

Figure 4: Fermi surface differences from Fe compounds to Ni compounds, here for (a) BaFe$_2$As$_2$ from ref. [67] and (b) BaNi$_2$As$_2$ from ref [32] projected onto the ab-plane. Dashed line indicates the surface where the superconducting order parameter is expected to change sign in the s± pairing state. Note that in (a) it does not intersect the Fermi surface, and thus the system will be fully gapped even in the s± state, while in (b) gapless (nodal) excitations are expected.

Band structure calculations also indicate a Van-Hove singularity in the density of states at an intermediate doping level between the Fe- and Ni-end members, which by a simple Stoner criteria suggests a ferromagnetic instability. Co lies between Fe and Ni on the periodic table, and indeed, LaCoAsO and LaCoPO are ferromagnets (T$_C$ = 66 K and 43 K respectively)[68,69], while

BaFeNiAs$_2$ and BaCo$_2$As$_2$ are suggested to lie in close proximity to a FM quantum critical point[43,70] (see figure 3b). The Sommerfeld coefficient of the latter two compounds is more than a factor of 2 larger than either BaFe$_{1.4}$Ni$_{0.6}$As$_2$ or BaNi$_2$As$_2$, consistent with the enhanced density of states from the Van-Hove singularity.

The phase diagram in figure 3b also highlights the two biggest differences already alluded to between the Fe-based systems and the Ni-based systems. Namely, in the Fe-based systems the parent compounds possess antiferromagnetism, which is suppressed with doping or pressure giving way to a dome of superconductivity. In the Ni compounds the parent compounds themselves superconduct, and there is no evidence to date for any proximity to magnetism (aside from the enhanced Wilson ratio in La$_3$Ni$_4$P$_4$O$_2$[8]), despite early calculations which suggested that they also lie close to a magnetic instability[36]. The other big difference is the roughly factor of 20 difference between the T$_c$ of the doped Fe system to that of the Ni analog. In fact, this is the basis on which band structure calculations claim that the Fe-systems must be unconventional, while the Ni ones may not be. Not surprisingly, calculations of the phonon spectra and the $\alpha^2F(\omega)$ function provide sufficient glue ($\lambda$ = 0.58 and 0.76 for LaNiPO and BaNi$_2$As$_2$, respectively) to produce transition temperatures up to 4 K, sufficiently above the observed transition temperatures[28,32]. However, one cannot account for the magnitude of T$_c$ in the Fe-based systems via phonon mediated pairing with similar calculations[71,72].

Surprisingly the renormalization effects on the electronic structure between the Fe- and Ni-end members appear to be similar. dHvA results on AFe$_2$As$_2$[73], LaFePO[74], and BaNi$_2$P$_2$[63] all suggest a mass renormalization m$_{exp}$/m$_{band}$ ~ 2. In the case of the Fe compounds, this result is supported by ARPES measurements[75]. Heat capacity measurements also generally support a mass renormalization of 2 in most cases for both systems (see table 1 for the Ni-compounds).

Whether the experimentally measured mass renormalization is predominantly due to phonons or contains additional correlation effects is currently unknown. It would be interesting to see if calculations including correlation effects such as DMFT would also support a correlated state in the Ni-based compounds as it has for the Fe-based systems[76]. It should be noted that the theoretically stable position of the pnictide atom relative to the Ni plane (the z parameter) agrees with the experimentally measured position. This supports the experimental results that magnetism and/or spin fluctuations do not appear relevant for the Ni-based compounds.

**Structure-Property relationships**

An empirical method for finding superconductors with higher $T_c$, is to identify structure-property relations that correlate with $T_c$. With this in mind, we list several structural parameters of the Ni-based superconductors in Table 2. So far, clear structure-$T_c$ relationships have been difficult to identify in the FeAs systems. Perhaps the best correlation to date has been in the ZrCuSiAs structure type, where $T_c$ appears to be a smoothly varying function of the As-Fe-As bond angle, with the optimum $T_c$ obtained close to the ideal tetrahedral angle of 109.5°[77]. It is noted that all of the Ni-based systems have an angle above 117° and are thus consistent with smaller $T_c$s, The exception to this are the borocarbides (discussed below), which are strikingly much closer to the ideal tetrahedral angle.

Table 2. Structural parameters of Ni based SC's in order of descending transition temperatures. c' (and V') is the distance (and corresponding volume) between neighboring $Ni_2X_2$ planes.

| | a (Å) | c' (Å) | V'(Å$^3$) | c'/a | $d_{Ni-Ni}$ (Å) | $d_{Ni-X}$ (Å) | $\angle_{XNiX}$ | refs |
|---|---|---|---|---|---|---|---|---|
| $LuNi_2B_2C$ | 3.464 | 5.316 | 63.78 | 1.53 | 2.449 | 2.102 | 110.94 | 78 |
| LaNiPO | 4.0453 | 8.1054 | 132.6 | 2.004 | 2.86 | 2.265 | 126.5 | 3 |
| GdNiBiO | | | | | | | | |
| $LaNiBiO_{1-x}$ | 4.073 | 9.301 | 154.3 | 2.284 | 2.880 | | | 16 |
| LaNiAsO | 4.1231 | 8.1885 | 139.2 | 1.986 | 2.915 | 2.3463 | 122.95 | 11 |
| $BaNi_2P_2$ | 3.947 | 5.91 | 92.071 | 1.497 | 2.791 | 2.259 | 121.7 | 51 |
| $La_3Ni_4P_4O_2$ | 4.0107 | 8.232, 4.858 | 132.4, 78.2 | 2.052, 1.211 | 2.836 | 2.271, 2.306 | 120.8, 124.0 | 8 |
| $SrNi_2P_2$ | 3.951$^b$ | 5.216$^b$ | 81.42$^b$ | 1.320$^b$ | 2.800$^b$ | 2.247-2.299 | 117.82-123.68 | 51 |
| $La_3Ni_4Si_4$ | 4.131, 4.176$^a$ | 11.79 | 203.4 | 2.839 | 2.938 | 2.427, 2.372 | 118.68, 121.06 | 79 |
| $La_3Ni_4Ge_4$ | 4.202, 4.217$^a$ | 12.02 | 213.0 | 2.85 | 2.989 | 2.447, 2.421 | 119.55, 121.46 | 79 |
| $BaNi_2As_2$ | 4.142 | 5.825 | 99.9 | 1.406 | 2.929 | 2.405 | 118.86 | 80 |
| $SrNi_2As_2$ | 4.154 | 5.145 | 88.78 | 1.239 | 2.937 | 2.377 | 121.82 | 81 |

$^a$ = a and b lattice parameters for the orthorhombic structure
$^b$ = averaged parameters for low T structure

It is remarkable that despite the huge difference in $T_c$, several trends of $T_c$ across different compounds appear to be similar in Ni- and Fe-based systems. For example, $T_c$ in the LaNiAsO system increases with both hole and electron doping, and roughly to the same value. While the parent compound LaFeAsO is not superconducting, both hole and electron doping produces very similar behavior to the Ni-compound as illustrated in figure 3a. It is also noted that in going from $LaTXO$ to $BaT_2X_2$ to $SrT_2X_2$ to $CaT_2X_2$ a monotonic suppression in $T_c$ is observed whether (T,X) = hole doped (Fe,As), (T,X) = (Ni,As), or (T,X) = (Ni,P), which can be seen in the columns of table 3. One possible origin for this trend is that dimensionality is reduced across this series. Arguments for reduced dimensionality enhancing $T_c$ in the Fe-based materials have already been made[82], and are consistent with specific theoretical models[83] and empirical trends in other material classes[84,85]. In this review we have suggested similar rationales to understand several observations in the Ni compounds (specifically $La_3Ni_4P_4O_2$ and $SrNi_2P_2$[10]). To understand the dimensionality between compounds mentioned above we note that the $ThCr_2Si_2$ structure is more 3D than the ZrCuSiAs structure, and one would anticipate increased 3-dimensionality in going from Ba to Sr to Ca as the smaller ionic size allows the $T_2X_2$ planes to increase hybridization along the c-axis relative to the ab-plane, as evidenced by the shrinking c/a ratio. To see if this hypothesis of increased dimensionality is indeed the dominant factor in reducing $T_c$ from LaTXO to $BaT_2X_2$ to $SrT_2X_2$ to $CaT_2X_2$, requires more detailed study into the electronic structure of these compounds. These observations also suggest that the synthesis of SmNiPO or ANiPF (A=Ca, Sr, Ba) may produce the highest $T_c$ of any Ni-based system to date.

Table 3: Comparison of superconducting transition temperatures across families in the ZrCuSiAs and $ThCr_2Si_2$ structure. Note the monotonically decreasing dependence of $T_c$ in going either across a row or down a column. Values for the (T = Fe, X = As) column are obtained from doped samples. Values from refs [5,39,59,86] and table 1.

|  | T=Fe, X=As | T=Ni,X=P | T=Ni,X=As |
| --- | --- | --- | --- |
| LaTXO | 43 K | 4.3 K | 2.75 K |
| $BaT_2X_2$ | 38 K | 2.7 K | 0.68 K |
| $SrT_2X_2$ | 37 K | 1.4 K | 0.62 K |
| $CaT_2X_2$ | 20 K |  |  |

Another perspective looking at table 3 would argue that $T_c$ is monotonically suppressed in going across rows from $Fe_2As_2$ to $Ni_2P_2$ to $Ni_2As_2$ planes irrespective of the host structure. To understand this, it may be important to recognize that conventional superconductivity is a low energy theory, for which the electronic structure above the cutoff frequency of the bosonic spectrum (typically chosen to be the phonon Debye frequency) is irrelevant. Band structure calculations show that the electronic structure near the Fermi energy is composed primarily of states from the $T_2X_2$ planes for both Ni-based and Fe-based systems. Consequently, aside from fine tuning of the electronic and magnetic structure it is perhaps not surprising that variations in the structure outside the $T_2X_2$ planes have relatively smaller effects on $T_c$ (aside from possibly tuning the dimensionality) than if the atoms within the $T_2X_2$ planes are substituted. In other words, $T_c$ which is determined by the bosonic spectrum, the quasiparticle spectrum, the coupling between them, and the electronic correlations is essentially set by the atoms of the $T_2X_2$ planes and just so happen to conspire such that it is an intrinsic property that electron pairs condense more easily in $Fe_2As_2$ layers, versus $Ni_2P_2$ layers, versus $Ni_2As_2$ layers. Thus, when comparing across these series it is a relatively minor effect whether layers of LaO, Ba, Sr, or Ca is used as the structural glue to hold the $T_2X_2$ layers together.

**Implications for similar and different pairing mechanisms**.

If the Ni compounds do indeed have the same pairing mechanism as the Fe compounds, but simply are not as well optimized for superconductivity, the fully gapped nature of the $BaNi_2As_2$ implies that the s± state would be exceedingly difficult to realize. The reason is that due to the more complex Fermi surface in the $BaNi_2As_2$ system, it is much more difficult to find a nodal plane which does not intersect the Fermi surface in $BaNi_2As_2$. Hence, the s± state is not supported by the results on the nickel compounds to date.

On the other hand, even if the pairing mechanisms are different, one can ask about the many empirical similarities that still exist between the Fe-based systems and the Ni-based systems. Perhaps, this is simply a statement of the notion that some crystal structures just happen to be good for superconductivity. For example the Perovskite structure supports d-wave $((La,Ba)_2CuO_4)$[19], p-wave $(Sr_2RuO_4)$[20], and s-wave $((Ba,K)BiO_3)$[21] pairing symmetries, while the $ThCr_2Si_2$ structure, in which all the transition-metal pnictide superconductors share a common

$T_2X_2$ structural element, is already known to support spin-mediated (ie $CePd_2Si_2$), valence-fluctuation-mediated ($CeCu_2(Si,Ge)_2$)[87], and phonon-mediated (ie. $LaPd_2Ge_2$[88]) superconductivity. There is no clear reason why this should be the case. The great variations in the electronic structure, bosonic fluctuation spectrum, not to mention the coupling between them and correlation effects within a given crystal structure seems at odds with the notion that a particular crystal structure would generally be favorable to superconductivity. However, perhaps this empirical observation hints at a deeper relation between crystal structure and the physical properties such as screening or electron-boson coupling which control superconductivity.

### Relationship to the borocarbides

We briefly discuss the nickel borocarbides and boronitrides, as they also contain $Ni_2X_2$ (X=B) planes as shown in figure 1g which would suggest that they may be related to the other Ni-based superconductors we have discussed above. The borocarbide systems have relatively high $T_c$s (ie. 16.5 K for $LuNi_2B_2C$[22]), and there is no magnetism associated with the Ni atoms (magnetism when present is due to the rare earths atoms in the structure, similar to REFeAsO or $EuNi_2As_2$). One potentially significant difference is that the ionic nature of the $(T_2Pn_2)^{2-}$ (T = Fe, Ni; Pn = pnictide) is presumably altered for the $Ni_2B_2$ planes. The pairing symmetry is still unresolved with reports both for and against unconventional superconductivity[89]. Future work will hopefully, help clarify the relationship between the structurally similar borocarbides and boronitrides, to the $Ni_2X_2$ (X=P, As, Bi, Si, Ge) systems. It would certainly also be interesting if a Fe, Ru, or Os borocarbide superconductor could be found.

**Conclusions and open questions**:

Clearly, more work is needed to elucidate the relationship between Fe-based and Ni-based pnictides. The fact that so many Ni analogs of the Fe superconductors also superconduct and vice versa suggests several interesting materials to attempt to synthesize including $La_3Fe_4P_4O_2$, $AFe_2P_2$ (A = Ba, Sr, Ca), Fe-, Ru-, or Os- based borocarbides, LiNiPn, SrNiPnF(Pn = pnictide), and $La_3Fe_4X_4$ (X = chalcogenide) to name a few. Direct comparisons of similar techniques on both the Fe- and Ni-based systems, would be particularly helpful to determine the relationship between the two families of superconductors. This suggests performing NMR, ARPES, penetration depth, and DMFT calculations on the Ni-systems.

In conclusion, we have shown that the stoichiometric compounds containing $Ni_2X_2$ (X = pnictide or chalcogenide) planes when examined outside the context of the FeAs superconductors appear to be conventional BCS phonon-mediated superconductors. The strongest evidence for which is from field dependent thermal conductivity measurements on $BaNi_2As_2$[31]. However, the increase of $T_c$ with doping in LaNiAsO, the similar mass renormalization to the Fe-analogs in the absence of magnetism, and the similar structure-property relations between the two remain to be understood.

**Acknowledgements**:

We thank our collaborators R.J. Cava, H. Lee, H. Nakote, A. Schultz, B.L. Scott, D.J. Singh, A. Subedi, and H. Volz, for experimental assistance and fruitful discussions. We also thank D.J. Singh for permission to create figure 4. Work at Los Alamos was performed under the